\documentclass[twocolumn,showpacs,showkeys,amsmath,amssymb,superscriptaddress,floatfix,nofootinbib]{revtex4}

\usepackage{graphicx}
\usepackage{bm} 
\usepackage{subfigure}

\def\vec#1{\mathchoice{\mbox{\boldmath$\displaystyle#1$}}
{\mbox{\boldmath$\textstyle#1$}}
{\mbox{\boldmath$\scriptstyle#1$}}
{\mbox{\boldmath$\scriptscriptstyle#1$}}}
\makeatletter

\begin{document}
\begin{titlepage}

\large{EUROPEAN ORGANISATION FOR NUCLEAR RESEARCH (CERN)}

\vspace*{1.5cm}

\begin{flushright}
{\small CERN-PH-EP/2009-016} \\
{\small 27 August 2009} \\
\end{flushright}
 
\title{Femtoscopic signatures of collective behavior as a probe of the
thermal nature of relativistic heavy ion collisions%
\footnote{Supported by the U.S. NSF Grant No. PHY-0653432.}}

\author{Adam Kisiel} 
\email{kisiel@if.pw.edu.pl}
\affiliation{Physics Department, CERN, CH-1211, Gen\`eve 23, Switzerland}
\affiliation{Faculty of Physics, Warsaw University of Technology, PL-00661 Warsaw, Poland}
\author{T. J. Humanic}
\email[]{humanic@mps.ohio-state.edu}
\affiliation{Department of Physics, The Ohio State University,
Columbus, OH 43210 USA}


\begin{abstract}
Femtoscopy measures space-time characteristics of the particle
emitting source created in relativistic heavy-ion collisions. It is
argued that collective behavior of matter (radial flow) produces
specific femtoscopic signatures. The one that is best known, the $m_T$
dependence of the pion ``HBT radii'', can be explained by the
alternative scenario of temperature gradients in an initial state
thermal model. We identify others that can invalidate such
alternatives, such as non-identical particle correlations and $m_T$
scaling for particles of higher mass. Studies with a simple
rescattering code show that as the interaction cross-section is
increased the system develops collective behavior and becomes more
thermalized at the same time, the two effects being the natural
consequence of increased number of particle rescatterings. Repeating
calculations with a more realistic rescattering model confirmed all of
these conclusions and provided deeper insight into the mechanisms of
collectivity buildup, showing a preference for a thermal model with
uniform temperature.  

\end{abstract}

\pacs{25.75.-q, 25.75.Dw, 25.75.Ld}

\keywords{relativistic heavy-ion collisions, hydrodynamics,
femtoscopy, non-identical particle correlations, elastic rescattering}

\maketitle 
\end{titlepage}


\section{Introduction}

The collective behavior of matter has been of central interest at the
Relativistic Heavy Ion Collider (RHIC), and will be an important part
of the research program at the Large Hadron Collider (LHC). It has
been studied through momentum observables: the inclusive $p_{T}$
spectra~\cite{Abelev:2008ez}, the elliptic flow coefficient
$v_2$~\cite{Voloshin:2008dg} as well as through space-time ones, such
as the $m_{T}$ dependence~\cite{Adams:2004yc} and azimuthal
oscillations of the femtoscopic ``HBT radii''~\cite{Adams:2003ra} and
the emission  asymmetries between non-identical particle
species~\cite{Adams:2003qa}. Hydrodynamic 
models have been able to describe the momentum observables with a
broad range of initial conditions and equations-of-state
(EOS)~\cite{Hirano:2002ds,Zschiesche:2001dx,Heinz:2002un}. Only 
recently it was shown that the space-time observables can be described
in the same framework only if particular choices about the initial
state and  the equation of state are made and resonance contributions
are fully taken into
account~\cite{Broniowski:2008vp,Pratt:2008qv}. However applying
hydrodynamic equations  
implies a strong assumption that the system is in local thermal 
equilibrium. Therefore studies were made~\cite{Gombeaud:2009fk} on how
relaxing this assumption would change this behavior. In particular the
study of elliptic flow $v_2$ data has led to the conclusion that the
Knudsen number (connected to the mean free path of a particle or the
interaction cross-section relative to the system size) does not reach
the hydrodynamic limit. Recently, influence of the Knudsen number (or
the assumed thermalization) on the femtoscopic observables has been 
investigated. 

The relaxation of the strong assumptions of hydrodynamics can also be
achieved by doing a microscopic rescattering simulation. The
hydrodynamic limit of parton cascades has been studied
before~\cite{Zhang:1998tj}, including the influence of the rescattering
cross-section on spectra~\cite{Cheng:2001dz} and freeze-out
patterns~\cite{Molnar:2002bz}. The amount
of interactions per particle can be controlled, by changing the ratio
of the interaction cross-section to the overall system size. By
adjusting the parameters so that the number of per-particle
rescatterings becomes large, one should approach the specific limiting
behavior of
hydrodynamics~\cite{Hirano:2002ds,Zschiesche:2001dx,Heinz:2002un},
which is also observed in blast-wave type
simulations~\cite{Retiere:2003kf,Kisiel:2006is}. 
In this work we aim to perform such
calculations by employing a simple model of elastic particle
rescatterings and compute the femtoscopic observables for the system
of pions and kaons. Two types of initial conditions are used, one with
and one without a temperature gradient in a thermal model. All other
parameters are kept fixed, so that a  
clean comparison between initial conditions and various values of the
cross-sections can be made and physics causes for observable
effects identified. In particular the $m_T$ dependence of the ``HBT
radii'' and the emission asymmetry between pions and kaons is studied
in detail. We focus on their dependence on the interaction
cross-section. The initial system size is kept fixed, so the
cross-section is directly correlated with average number of collisions
per particle and the Knudsen number. In addition, we use a
more realistic rescattering model which also includes nucleons, inelastic
scattering and low-lying resonances to cross check some our results
with the simple rescattering model.

The paper is organized as follows: In Section~\ref{sec:model} the
simplified model for particle rescattering is described. In
Section~\ref{sec:observabledef} we outline the procedure to obtain the
relevant observables. In Section~\ref{sec:results} the results
from the simple model are shown, and in Section~\ref{sec:realistic}
results from the more realistic rescattering model are discussed. 

\section{Simplified model for particle rescattering}
\label{sec:model}

The calculation is initialized by distributing a pre-defined number of
pions and kaons into a limited volume, according to the following
density profile: 
\begin{equation}
\frac {dN} {dx dy dz} \approx exp(- \frac{x^2 + y^2} {2 R^2}) \Theta(z_{max}
- z) \Theta(z_{max} + z) 
\end{equation}
where $R$ is the transverse Gaussian radius of the system, $z_{max}$
is its longitudinal extent and $\Theta$ is a step function. The
transverse momentum of the  particle is generated randomly from the
thermal distribution, with the temperature at a given point: 
\begin{equation}
T(x, y) = (T_{max} - T_{base}) exp(- \frac{x^2+y^2} {2 R_{T}^2}) + T_{base},
\end{equation}
while the velocity profile in the longitudinal direction is
semi Hubble-like: 
\begin{equation}
V_z = a * p_z + b,
\end{equation}
where $b$ is a random component. The model has six
parameters: the transverse source size $R$, maximum and minimum
temperature $T_{max}$ and $T_{base}$ and the parameter controlling its
gradient $R_{T}$, the longitudinal extent of the source $z_{max}$ and
its velocity scaling factor $a$. Out of these only the first four will
be important in this work, because we will focus on the transverse
plane.  We emphasize that the particular form of the initial state has
been specifically chosen so that it does not include any transverse
collective features. Any collectivity signals can only develop via
particle interactions. 

Once the particles are distributed in the initial state, the system 
evolves according to the following procedure. All particles are
combined into pairs and for each pair a possibility of a collision is
determined. The collision occurs if the time of the collision is
within the current time slice (taken as $0.1$~fm) and the
particles are moving towards each other, that is:
\begin{equation}
(\vec r_1 - \vec r_2)\cdot(\vec v_1 - \vec v_2) < 0,
\end{equation}
where $\vec r=(x,y,z)$ is the position of the particle and $\vec
v=\vec p/E$ is its velocity. Their distance at the point of closest
approach must be less than the collision distance 
$d$ (which is the crucial parameters that is varied for different
simulations). All such pairs are recorded. Then all the rescatterings
for a given time step are carried out at the same time, with each
particle possibly colliding more than once. Relativistic kinematics is
used and the scatterings are fully elastic. The  scattering
particles are replaced by the rescattered ones. Then the procedure is
repeated for the next time step. The evolution is carried out
until no scatterings remain. Then particles for the next ``event'' are
initialized and the rescattering procedure is repeated.

Finally, at the end of the evolution particles are written out as
``events'' i.e. collections 
of pions and kaons participating in the same rescattering
procedure. The emission point for each particle is taken as the point
of the last scattering and is saved along with the momentum
information. These events are later analyzed to extract 
observables mentioned below.

We emphasize that the aim of this simple model is not to accurately
describe the rescattering process taking place in the heavy-ion
collision. Instead we aim to determine, with as simple calculation as
possible, whether a microscopic type of rescattering simulation,
without strong assumptions of full thermalization  of matter can still
produce collective behavior of matter. In other words we test the
degree to which the conclusions from the hydrodynamic models can be
viewed as general ones.

Most parameters of the model are kept fixed to make the comparisons
easy. The system size $R$ is $4.0$~fm while the longitudinal extent 
is $2$~fm. Each ``event'' consists of 1000 particles with pion mass
and 100 particles with kaon mass. For each set of parameters 1000
events were simulated. Two cases are considered: in the ``uniform''
case the temperature is uniform in the whole system, that is $T_{base}
= T_{max} = 300$~MeV; in the ``gradient'' case $T_{max} =~500$~MeV and
$T_{base} = 100$~MeV. $R_{T} = 2$~fm. The whole study is repeated for
both initial configurations and as a function of the interaction
distance (and therefore the cross-section): $d = 0.1$~fm, $1$~fm,
$2$~fm, $5$~fm and $10$~fm. For the  given initial conditions this
corresponds to the average number of collisions per particle $\left <
N_{c} \right >$ of $0.15$, $2.2$, $4.7$, $9.2$ and $12.9$
respectively. 

\section{Definition of observables}
\label{sec:observabledef}

Since ``collectivity'' and its consequences for the thermal nature of
the system is the main focus of this
paper we begin by defining what we mean by that term. We consider a set
of particles which are close to each other in space-time. We calculate
their average velocity $\left < \vec v \right >$. If one observes this
velocity to be non-zero, in a consistent manner, for many cells
located thorough the system, we consider it to behave
collectively. One immediately notices a similarity to hydrodynamics, 
where one considers a fluid element localized in space-time and its
velocity 
taken from the flow field. However in our case we do not require that
particles are locally thermalized in a cell, only that they have a
common velocity.

This general definition allows the flow velocity $\vec v$ to have any
direction, but the system we consider has radial symmetry in the
transverse plane, with large density center and low density exterior,
so we expect the flow to develop in the ``outwards'' direction -
pointing outside from the center of the source. Therefore to directly
probe collectivity one should look at the average ``outwards''
velocity of particles: 
\begin{equation}
\left < v_{out} \right > = \frac {\left < \vec v_T \cdot \vec r_T \right >} {\left | \vec r_{T} \right |},
\label{voutdef}
\end{equation}
where the $T$ subscript refers to the $(x,y)$ transverse plane.
If one observes a $\left < v_{out} \right >$ which is non-zero and
positive we declare such system to have ``radial flow''. In contrast
one expect the average of the other-''sideward'' component of the
transverse velocity to be zero:
\begin{equation}
\left < v_{side} \right > = \frac {\left < \vec v_T \vec \times \vec r_T \right >} {\left | \vec r_{T} \right |} =0.
\label{voutdef}
\end{equation}
We also expect that the as the cross-section grows, the system moves
towards a hydrodynamic limit - that is the dependence of $\left <
v_{out} \right >$ vs. transverse radius $r_T$ is monotonously and
smoothly growing and the collective velocity of particles of different
masses (here pions and kaons) is the same~\cite{Chojnacki:2006tv}. Note
that such plots can only be made in models, and do not correspond to
any experimental observable. They are however useful to confirm that
the system which we simulate does indeed exhibit ``collective'' behavior.

\begin{figure}[tb]
\begin{center}
\includegraphics[angle=0,width=0.48 \textwidth]{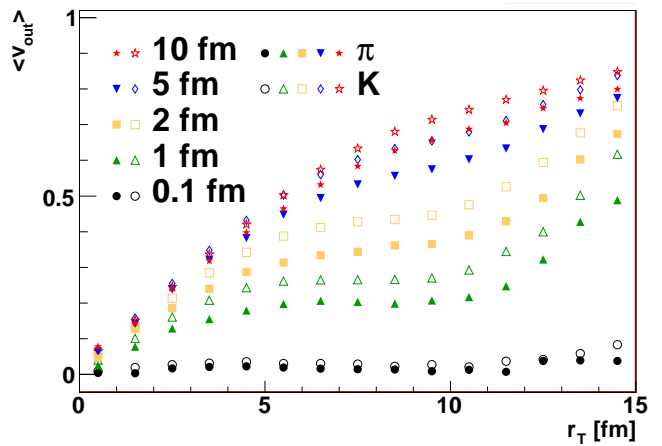}
\end{center}
\vspace{-6.5mm}
\caption{(Color on-line) Average collective velocity $\left < v_{out}
\right >$ for different values of interaction distance $d$. Black
circles are for $d = 0.1$~fm, green up triangles: $1$~fm, yellow
squares: $2$~fm, blue down triangles or diamonds: $5$~fm, red stars:
$10$~fm. Open symbols are pions, closed: kaons.
\label{fig:pikvels}}
\end{figure}

Moving to the femtoscopic analysis, one takes the events as input and
extracts the information in the following way: The ``outwards'' and
``sidewards'' distribution of the particles' emission points are
obtained as a function of particle's $p_T$ (or equivalently $m_T =
\sqrt{p_{T}^2 + m^2}$ or $v_T = p_T / m_T$):
\begin{eqnarray}
x_{out} = \frac {\vec r_T \cdot \vec v_T} {\left | \vec v_{T}
\right |} - v_T t \nonumber \\
x_{side} = \frac {\vec r_T \times \vec v_T } {\left | \vec v_{T}
\right |}  
\end{eqnarray}
Then each slice in $p_T$ ($m_T$, $v_T$) is fitted with a Gaussian. The
sigmas of this Gaussian $X_{out}$ or $X_{side}$ are plotted. The
mean of the distribution $\mu_{out}$ is also calculated. In
non-identical particle femtoscopy one correlates particles with
similar velocity. Therefore to obtain an estimate of the emission
asymmetry one subtracts the mean of the distribution for pions from
the one of kaons, in the same $v_t$ bin.   

The single particle sigmas $X_{out}$ and $X_{side}$ are what is
usually reported in model studies as ``HBT radii'', therefore we show
them for comparison. However femtoscopy by definition probes the
characteristics of the ${\it two-particle}$ emission function also
called the ``separation distribution''. The real, measurable ``HBT
radii'', for which we will use the symbols $R_{out}$ and $R_{side}$
(to distinguish them from the single-particle estimates), can be used
to infer the values of $X_{out}$ and $X_{side}$ only if certain
assumptions about the emission function are made. Such procedure is
often required for hydrodynamic model studies where only the analytic
form of the emission function is known. It has been shown to introduce
a systematic uncertainty if they are compared to the
experimental HBT radii~\cite{Frodermann:2006sp}. In our case, since we
deal with particles, we can use the direct two-particle method to
calculate the actual 3D correlation function, and infer the values of
the ``HBT radii'' $R_{out},R_{side},R_{long}$ in the exact same manner
as the experiments do: by fitting them with the usual formula:
\begin{equation}
C(\vec q) = 1 + \lambda \left (- q_{out}^2 R_{out}^2- q_{side}^2
R_{side}^2- q_{long}^2 R_{long}^2 \right ),
\label{cffit}
\end{equation}
where $q$ is the relative momentum of the pair calculated in the
Longitudinally Co-Moving Frame (LCMS) (where the 
longitudinal pair velocity vanishes). In this way most systematic
uncertainties in comparing model sigmas and experimental ``HBT
radii'' are removed. For the detailed description of the
``two-particle'' method to calculate the CF, which is beyond the scope
of this work, we refer the reader to~\cite{Kisiel:2006is}.

\section{Analysis results}
\label{sec:results}

\begin{figure}[tb]
\begin{center}
\includegraphics[angle=0,width=0.48 \textwidth]{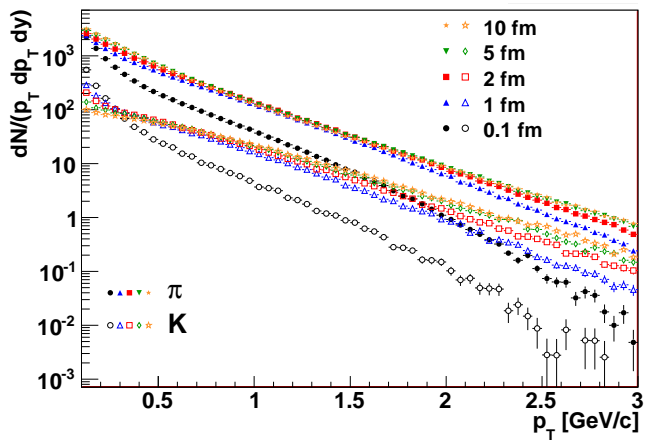}
\end{center}
\vspace{-6.5mm}
\caption{(Color on-line) Evolution of $p_T$ spectra of pions (closed
symbols) and kaons (open symbols) as a function of $d$. Black circles
are for $d = 0.1$~fm, blue up triangles: $1$~fm, red squares: $2$~fm,
green down triangles or diamonds: $5$~fm, orange stars: $10$~fm.
\label{fig:pikspectra}}
\end{figure}

\begin{figure*}[tb]
\begin{center}
\includegraphics[angle=0,width=0.9 \textwidth]{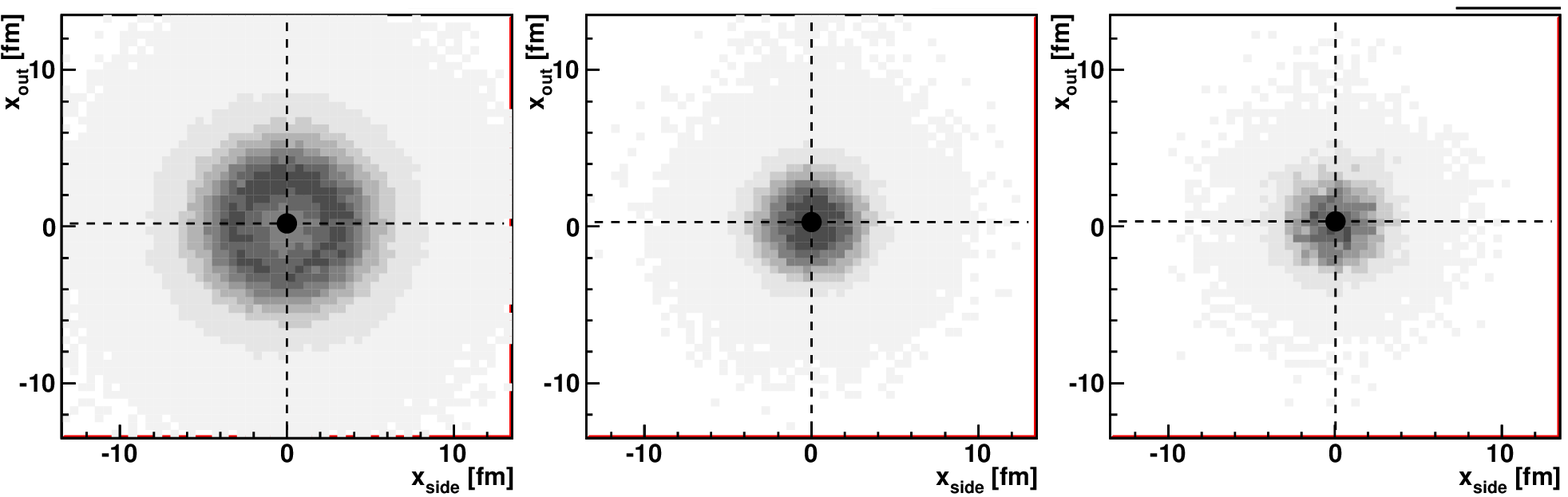}
\includegraphics[angle=0,width=0.9 \textwidth]{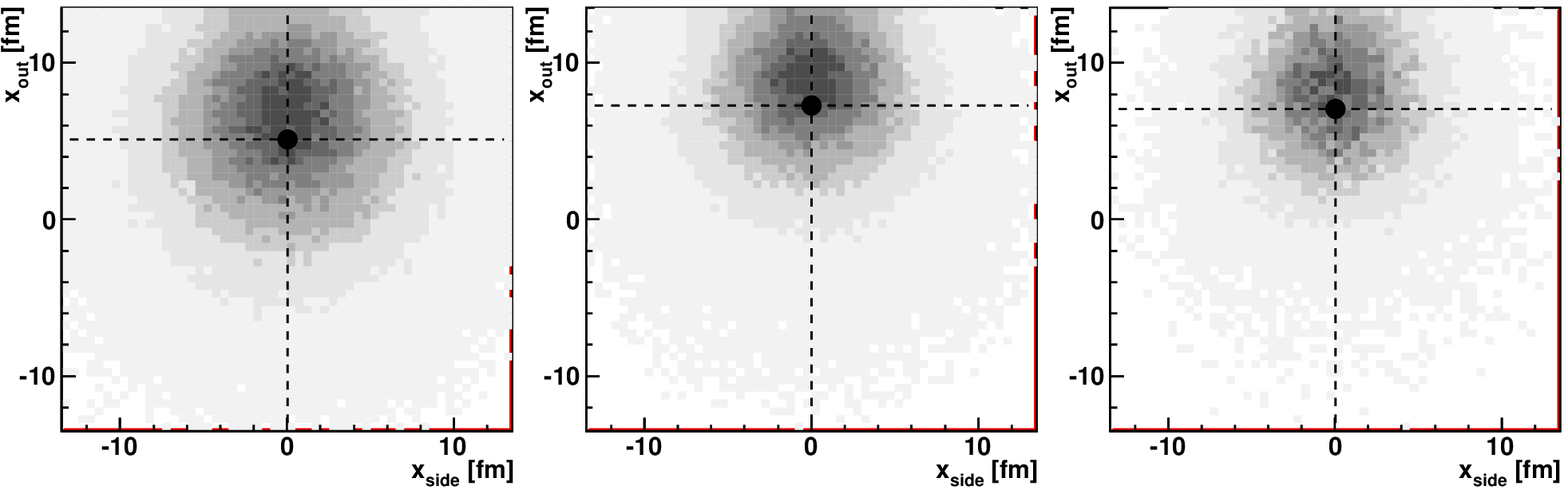}
\end{center}
\vspace{-6.5mm}
\caption{Distribution of emission points of pions (left panels, $0.15
< p_T < 0.25$~GeV/c, middle panels $0.4 < p_T < 0.6$~GeV/c) and kaons
(right panels $0.53 < p_T < 0.88$~GeV/c) for simulation with small
cross-section (upper panels $d = 0.1$~fm) and large cross-section
(lower panels $d = 10$~fm). Markers show average emission points. 
\label{fig:piksources}}
\end{figure*}

We begin our analysis by establishing whether we do observe
collectivity in the system produced by our calculations. The flow
velocity of particles is shown in Fig.~\ref{fig:pikvels}. We see that
for small interaction distance $d$ the system shows no
collectivity. As $d$ (and $\left < N_c \right >$) increases,
the collectivity develops, and has non-trivial shape vs $r_T$. 
For $d=10$~fm the shape of the flow profile already resembles the ones
typically produced by hydrodynamics
codes~\cite{Chojnacki:2006tv}. Also the kaon velocity is 
not exactly the same as pions, but it gets closer with increasing
$d$, again confirming our expectation that by increasing $d$ we
approach the limit of ideal hydrodynamic behavior.

In Fig.~\ref{fig:pikspectra} the the most basic observables - the
$p_{T}$ spectra, are shown for all values of $d$ for the
``gradient'' initial configurations. The original 
distribution is manifestly non-thermal-like for pions and kaons. As
the number of collisions grows, the distributions develop towards the
shapes well known from hydrodynamic calculations: thermal curves
modified by collective flow. Pions become almost exponential with
negative curvature, kaons develop positive curvature. So by increasing
the number of interactions the spectra develop signatures of a 
thermalized system with collective velocity. 

Next, let us visualize how does the emitting region look like as a
function of particle type, $p_T$ and $d$ used in the calculation. It
is shown in Fig.~\ref{fig:piksources}. Emission point coordinates of
each particle are projected on particle's velocity direction and 
plotted, so we obtain the emission picture in the relevant
``out-side'' coordinates. Upper plots show the behavior of the
``gradient'' scenario with small cross-section. The size of the emitting
region decreases with particle $p_T$, but it is not shifted - the mean
emission point stays close to zero. The picture is notably different
for the lower plots, where $d$ is large, resulting in many
interactions. Again one sees the decrease of the size of the emitting
region with $p_T$, but also a strong shift of the mean emission point in
the ``out'' direction. This shift also grows with the particle's
$p_T$. Both effects are natural consequences of radial flow and have
been observed in hydrodynamic
calculations~\cite{Hirano:2002ds,Zschiesche:2001dx,Heinz:2002un,Retiere:2003kf}.
This section is devoted to the discussion on how they are reflected in
the femtoscopic observables. 

\begin{figure*}[tb]
\begin{center}
\includegraphics[angle=0,width=0.9 \textwidth]{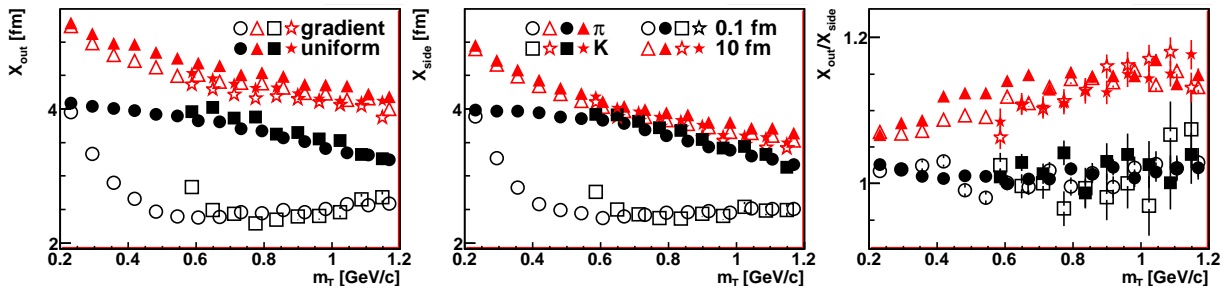}
\end{center}
\vspace{-6.5mm}
\caption{(Color on-line) Fitted sigmas of single particle distributions
as a function of $m_{T}$. Left panel shows outwards sigma, middle -
``side'', right - ``out'' over ``side'' ratio. Open symbols correspond
to ``gradient'', closed to ``uniform'' scenarios. Circles and squares
are for $d = 1$~fm, triangles and stars for $d = 10$~fm. Circles and
triangle are for pions, squares and stars for kaons. 
\label{fig:pikvars}}
\end{figure*}

\begin{figure}[tb]
\begin{center}
\includegraphics[angle=0,width=0.4 \textwidth]{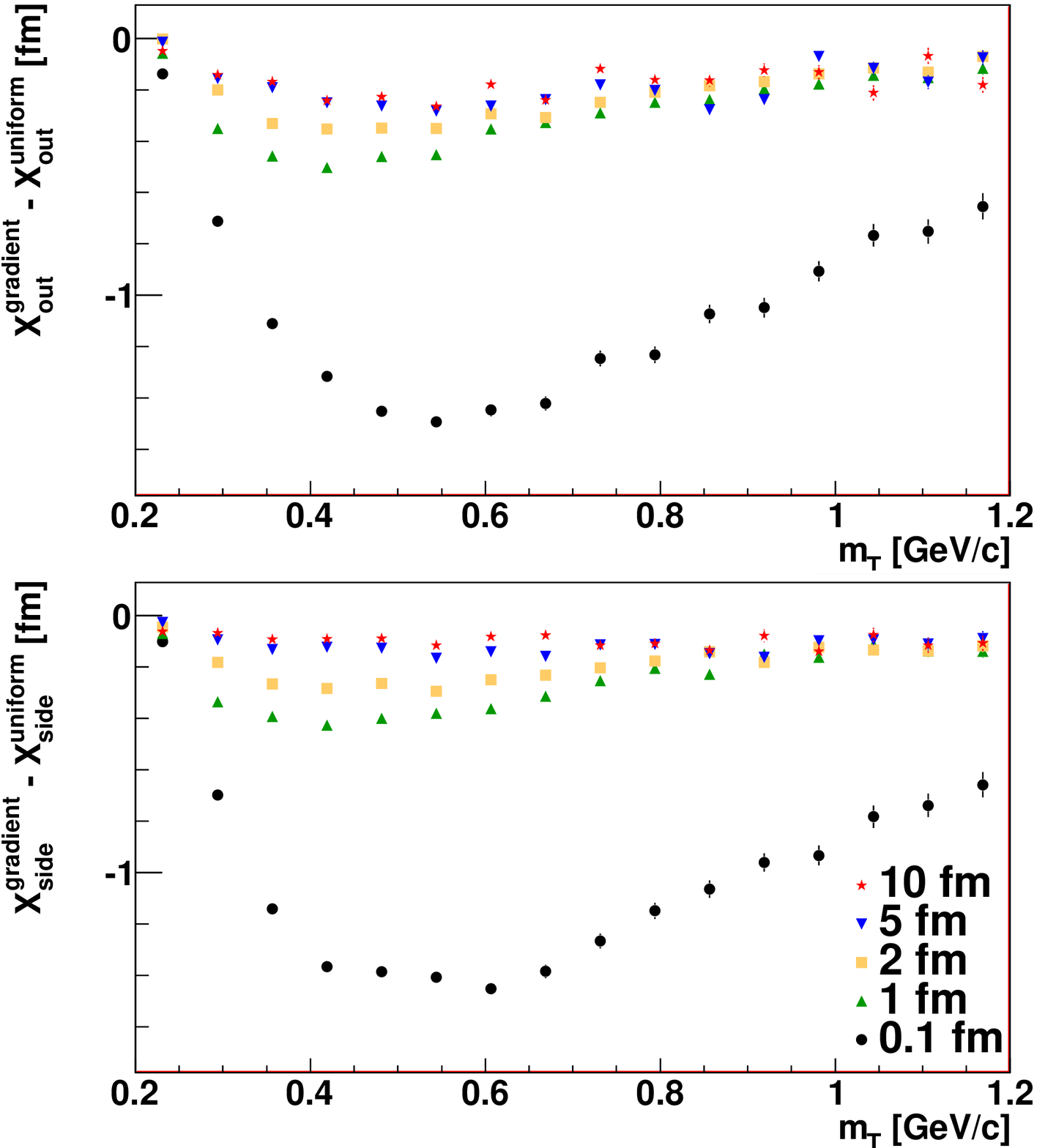}
\end{center}
\vspace{-6.5mm} 
\caption{(Color on-line) Differences in single particle sigmas
between ``uniform'' and ``gradient'' scenarios as a function of
$m_{T}$. Upper panel shows ``outwards'' difference, lower -
``sidewards''. Black dots: $d = 0.1$~fm, green up-triangles: $d =
1$~fm, yellow squares: $d = 2$~fm, blue down-triangles: $d = 5$~fm,
red stars: $d = 10$~fm.   
\label{fig:vardiffs}}
\end{figure}

In Fig.~\ref{fig:pikvars} the single particle sigmas $X_{out}$ in the
``out'' and $X_{side}$ in the ``side'' direction are compared, for
pions and kaons, ``uniform'' and 
``gradient'' scenarios and two extreme values of $d$. Let us analyze
the ``side'' sigma, which should reflect 
the effects of flow and temperature gradients with no additional
complication from emission duration. In the case of low cross-section
one can see small $m_T$ gradient for radii when no initial temperature
gradient is present. Introducing a temperature gradient makes the $m_T$
dependence steeper, although it flattens at large $m_T$. Also the
sigmas for kaons do not follow the universal $m_T$ scaling curve,
at least at low kaon $p_T$. The ``out/side'' ratio is almost flat at
1.05. Introducing rescattering has several effects. First of all the
system size grows, as the particles tend to rescatter longer. The
$m_T$ dependence becomes more pronounced than for small $d$ - a
signature of collective behavior of matter. Also sigmas of kaons
start to follow a universal $m_T$ scaling curve - another crucial
signature of collectivity. As one should expect, multiple scatterings
of particles lead to isotropization of the system - the information
about initial temperature gradients is lost - hence the difference
between ``gradient'' and ``uniform'' scenario becomes small and constant -
in other words both produce {\it the same} $m_T$ dependence of
radii. To emphasize the point, the evolution of the difference between
the two scenarios as a function of $d$ is shown in
Fig~\ref{fig:vardiffs}. One can see how the isotropization process
occurs gradually as the number of scatterings grows. This brings an
important conclusion: the $m_{T}$ dependence of ``HBT radii'' can be
caused by temperature gradients only in the case of very weak
interactions. In that case more massive particles do not follow
the $m_T$ scaling. Once the cross-section grows, the initial gradients
are forgotten and the expected signatures of collective behavior
develop: the universal $m_T$ dependence of sigmas for all
particles. 

\begin{figure}[tb]
\begin{center}
\includegraphics[angle=0,width=0.48 \textwidth]{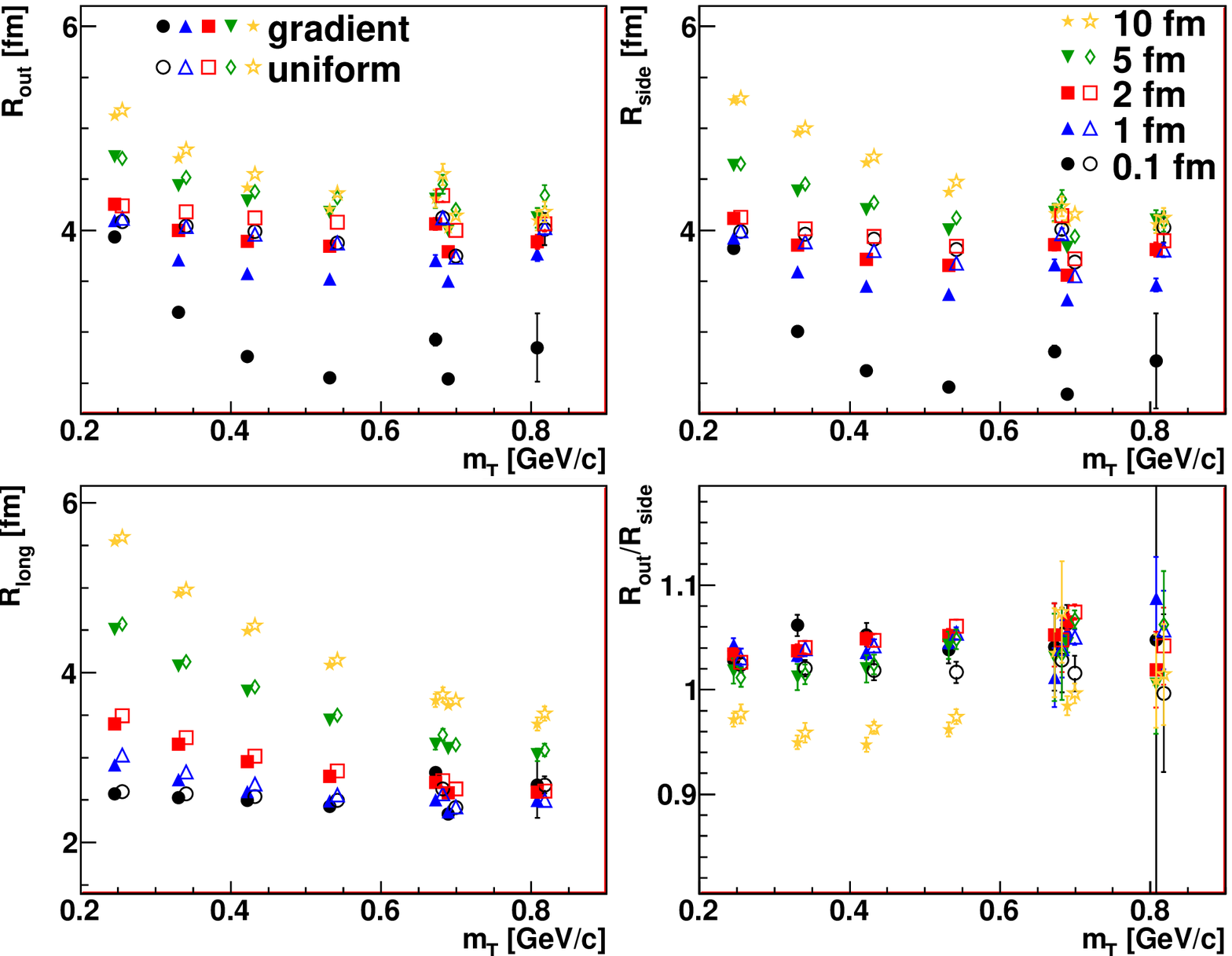}
\end{center}
\vspace{-6.5mm} 
\caption{(Color on-line) $m_T$ dependence of HBT radii. Upper-left
panel shows $R_{out}$, upper-right $R_{side}$, lower-left $R_{long}$,
lower-right $R_{out}/R_{side}$. Closed symbols are for ``gradient'',
open (shifted for clarity) for ``uniform'' initial conditions. Four
low $m_T$ points and the one second from the right are for pions, the
highest and third from the right in $m_T$ - for kaons. Circles,
up-triangles, squares, down-triangles and stars correspond to growing 
$d$ from $0.1$~fm to $10$~fm.  
\label{fig:hbtmtdep}}
\end{figure}

Similar dependence, seen in Fig.~\ref{fig:hbtmtdep} can be calculated
for the actual ``HBT radii'' obtained via the two-particle method. The
conclusions from the previous  
paragraph hold for real radii as well: at low cross-section $m_T$
dependence of radii can only come from temperature gradients, but $m_T$
scaling for heavier particles is violated. As cross-section grows the
universal $m_T$ behavior is recovered for particles of different
masses and the differences in initial conditions are washed
out. Interestingly, the $R_{out}/R_{side}$ ratio, in contrast to single
particle sigmas, is not growing and even decreases at very high
cross-section. The cause for this discrepancy was not investigated;
however, it is the ``HBT radii'' that are actually measured by the
experiment, the sigmas are just an approximation.

We have shown that in simple rescattering calculations it is possible
to produce an $m_T$ dependence of HBT radii both by temperature
gradients and collective behavior resulting from many
rescatterings. We have identified a clear signature that differentiate
the ``gradient'' and ``collective'' scenario: a universal $m_T$
scaling for particles of different masses. However it is not clear how
this scaling is affected by the strongly decaying resonance
contributions to pions. Also the 3D kaon HBT radii are difficult to
measure with sufficient accuracy. It would be desirable to identify an
observable which would be directly sensitive to the collectivity
arising from multiple particle interactions. One such observable,
which has already been measured at RHIC, is the emission asymmetry
between particles of different masses. It is measured in non-identical
particle femtoscopy~\cite{Adams:2003qa}.

In short, non-identical particle femtoscopy measures a difference
between mean emission points of particles of the same
velocity~\cite{Lednicky:1995vk,Lednicky:2003fe}. If we 
analyze pion-kaon pairs, similar velocity necessarily means large
difference in momenta. It has been argued that collectivity, which (as
we and many others have shown) produces the $m_T$ dependence of
radii, also produces a shift in the mean emission point in the out
direction, which is illustrated in Fig.~\ref{fig:piksources}. The two
effects are intimately connected and depend on 
particle's $p_T$. Therefore a similar velocity pion and kaon, having
very different $p_T$, will also have different mean emission
points. We call this difference the emission asymmetry. In contrast,
the non-collective ``gradient'' scenario, even 
though it produces the $m_{T}$ dependence of radii, has zero asymmetry. 

\begin{figure}[tb]
\begin{center}
\includegraphics[angle=0,width=0.48 \textwidth]{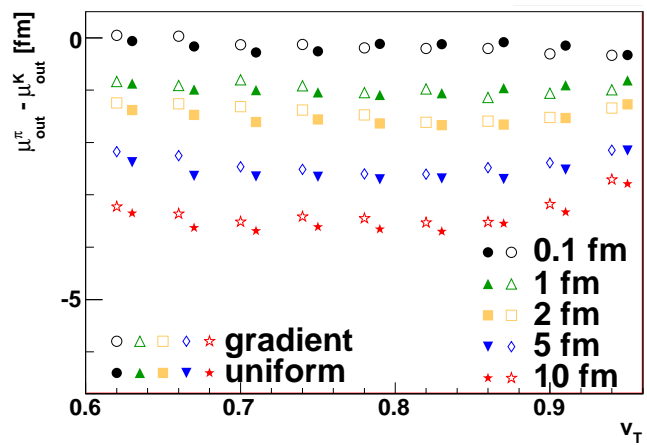}
\end{center}
\vspace{-6.5mm} 
\caption{(Color on-line) $v_T$ dependence of emission asymmetry between
pions and kaons. Closed symbols are for ``uniform'', open for
``gradient'' initial conditions. Circles, up-triangles, squares,
down-triangles and stars correspond to growing $d$, from $0.1$~fm to
$10$~fm.  
\label{fig:pikmeandiff}}
\end{figure}

\begin{figure}[tb]
\begin{center}
\includegraphics[angle=0,width=0.48 \textwidth]{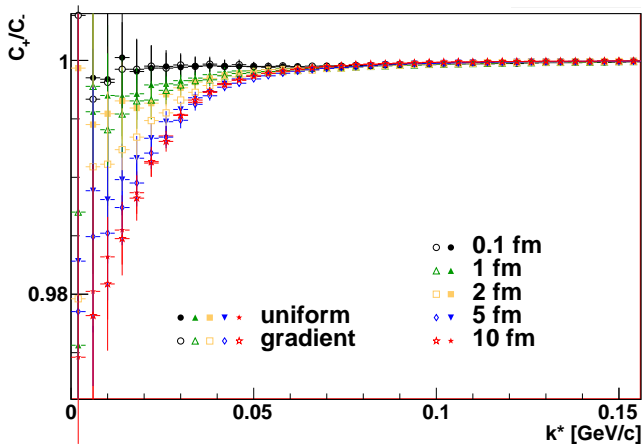}
\end{center}
\vspace{-6.5mm} 
\caption{(Color on-line) Experimental signature of emission asymmetry
between pions and kaons: the ``double ratio''. Closed symbols are for
``uniform'', open for ``gradient'' initial conditions. Circles,
up-triangles, squares, down-triangles and stars correspond to growing
$d$ from $0.1$~fm to $10$~fm. 
\label{fig:pikcfdr}}
\end{figure}

\begin{figure}[tb]
\begin{center}
\includegraphics[angle=0,width=0.48 \textwidth]{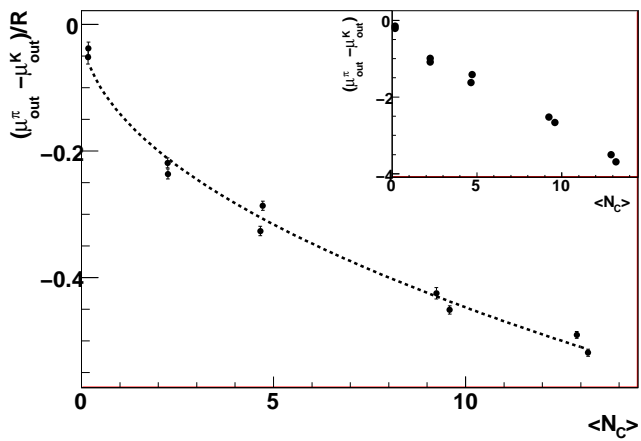}
\end{center}
\vspace{-6.5mm} 
\caption{Mean emission asymmetry between pions and kaons scaled by the
overall system size vs. average number of rescatterings per
particle. Line is a fit with $a\sqrt{b \left < N_C \right >}$. Inset:
Mean asymmetry vs. the average number of rescatterings.
\label{fig:pikshiftnc}}
\end{figure}

Fig.~\ref{fig:pikmeandiff} shows the difference between mean emission
points of pions and kaons as a function of particle's velocity, which
is the relevant variable for non-identical particle correlations. One
can see that small cross-section means small asymmetry, large
cross-section produces a significant one. Moreover, the asymmetry seems
to be, to a large degree, independent of the initial temperature
distribution; it is only affected by $d$. In other words
we have identified a good candidate for an observable that directly
probes the amount of collectivity in the system and can be used to
infer (in a model dependent way) the amount of interactions that
particles undergo which can be directly related to the Knudsen
number. In Fig~\ref{fig:pikcfdr} the measurable signature of the emission
asymmetry between pions and kaons - the non-identical particle
``double-ratio'' calculated with the two-particle method is
shown. Small cross-section results in the ``double ratio'' close to
unity (meaning no emission asymmetry), large cross-section shows
significant deviations from unity (meaning large asymmetry),
confirming that this experimental observable is sensitive to the 
amount of collectivity in the system. In fact in
Fig.~\ref{fig:pikshiftnc} we show that for our simple model the
asymmetry is directly proportional to the average number of
rescatterings per particle. Since our system has arbitrary initial
size we also show the asymmetry scaled by the overall system size, so
it can be compared to systems with other sizes.

The asymmetries probed by the non-identical particle correlations are
known to be influenced by the resonance decay products just as the
``HBT radii'' are. Fortunately, it was recently
demonstrated~\cite{Kisiel:2009zz} that decays, which are random in
their nature, cannot produce the  space-velocity direction
correlations which are naturally arising from 
collectivity. Moreover in the specific case of pions and kaons the
characteristics of the decay processes are such that they do
not dilute, but actually magnify the emission asymmetries coming from
``flow''. Therefore, even though a need for a more detailed and
realistic calculation is clear, non-identical particle correlations
should remain a clean way of probing the degree of collectivity in the
system 
produced in heavy-ion collisions. We also note that multiple
collisions per particle simultaneously produce collectivity (or 
space-velocity direction correlation) and system isotropization and,
in a longer timescale, thermalization. It would 
be difficult to come up with a scenario in which collectivity is not
correlated with thermalization. An extreme case of such correlation is
obviously the hydrodynamic model in which perfect thermalization and
complete collectivity is assumed.

\section{Realistic rescattering model}
\label{sec:realistic}

In order to check the validity of the simplified calculations above we
have repeated some of them using a more realistic rescattering model which
includes elastic and inelastic rescattering modes for pions, kaons, nucleons 
and low-lying resonances \cite{Humanic:2005ye}. A description of this
rescattering model now follows. Rescattering is simulated 
with a semi-classical Monte Carlo
calculation which assumes strong binary collisions between hadrons.
Relativistic kinematics is used throughout. The hadrons considered in the
calculation are the most common ones: pions, kaons,
nucleons and lambdas ($\pi$, K,
N, and $\Lambda$), and the $\rho$, $\omega$, $\eta$, ${\eta}'$,
$\phi$, $\Delta$, and $K^*$ resonances.
For simplicity, the
calculation is isospin averaged (e.g. no distinction is made among a
$\pi^{+}$, $\pi^0$, and $\pi^{-}$).

The rescattering calculation finishes
with the freeze out and decay of all particles. Starting from the
initial stage ($t=0$ fm/c), the positions of all particles in each event are
allowed to evolve in time in small time steps ($\Delta t=0.5$ fm/c)
according to their initial momenta. At each time step each particle
is checked to see a) if it has hadronized, b) if it
decays, and c) if it is sufficiently close to another particle to
scatter with it. Isospin-averaged s-wave and p-wave cross sections
for meson scattering are obtained from Prakash et al.\cite{Prakash:1993bt}
and other cross sections are estimated from fits to hadron scattering data
in the Review of Particle Physics\cite{Yao:2006px}. Both elastic and inelastic collisions are
included. The calculation is carried out to 50 fm/c or greater to
allow enough time for the rescattering to finish (as a test, calculations were also carried out for
longer times with no changes in the results). Note that when this cutoff time is reached, all un-decayed resonances are allowed to decay with their natural lifetimes and their projected decay positions and times are recorded.

Using this rescattering model we calculated
the two-particle separation distribution widths for pions and kaons comparable 
to Fig.~\ref{fig:hbtmtdep} and emission asymmetries comparable to
Fig.~\ref{fig:pikmeandiff}  for three initial state cases: 
1) the ``uniform'' and 2) ``gradient'' cases of the simple initial state model
defined by Eqs. (1)-(3), and 3) using an initial state model composed of a superposition of $p+p$ collisions as
was done in Ref.~\cite{Humanic:2008nt}.
It was shown in  Ref.~\cite{Humanic:2008nt} that when this model is coupled with the realistic rescattering model
described above, agreement with a wide
range of hadronic observables from RHIC experiments is obtained, including good agreement
with HBT measurements.

We present a brief description of the initial state model used in case 3). This model is based on
superposing PYTHIA-generated  $p+p$ collisions calculated at the beam $\sqrt s$
within the collision geometry of the colliding nuclei. 
The $p+p$ collisions were modeled with the PYTHIA code \cite{Sjostrand:2006za}, version 6.409.
Specifically, for a collision of impact
parameter $b$, if $f(b)$ is the fraction of the overlap volume of the participating parts of the nuclei such that $f(b=0)=1$ and $f(b=2R)=0$, where $R=1.2A^{1/3}$ and $A$ is the mass 
number of the nuclei, then
the number of $p+p$ collisions to be superposed will be $f(b)A$. The positions of the superposed $p+p$ pairs are randomly distributed in the overlap volume and then projected onto the $x-y$ plane which is transverse to the beam axis defined in the $z$-direction. The coordinates for a particular
$p+p$ pair are defined as $x_{pp}$, $y_{pp}$, and $z_{pp} = 0$. 
The positions of the hadrons produced in 
one of these $p+p$ collisions are defined with respect to the position so obtained of the superposed
$p+p$ collision (see later). 

The space-time geometry picture for hadronization from a superposed $p+p$
collision located at $(x_{pp},y_{pp})$ consists of the emission of a PYTHIA
particle from a thin uniform disk of radius 1 fm in the $x-y$ plane followed by
its hadronization which occurs in the proper time of the particle, $\tau$. The space-time
coordinates at hadronization in the lab frame $(x_h, y_h, z_h, t_h)$ for a particle with momentum
coordinates $(p_x, p_y, p_z)$, energy $E$, rest mass $m_0$, and transverse disk
coordinates $(x_0, y_0)$, which are chosen randomly on the disk,  can then be written as

\begin{eqnarray}
x_h = x_{pp} + x_0 + \tau \frac{p_x}{m_0} \\
y_h = y_{pp} + y_0 + \tau \frac{p_y}{m_0} \\
z_h = \tau \frac{p_z}{m_0} \\
t_h = \tau \frac{E}{m_0}
\end{eqnarray}

\begin{figure}[tb]
\begin{center}
\includegraphics[angle=0,width=0.48 \textwidth]{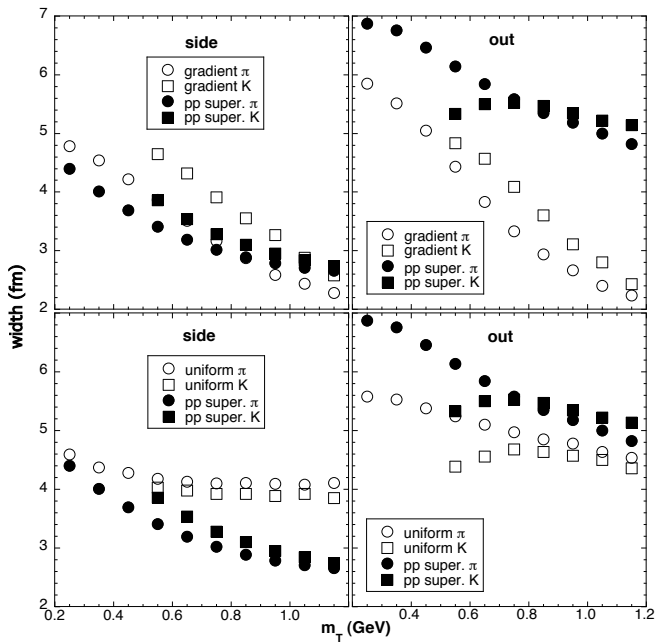}
\end{center}
\vspace{-6.5mm} 
\caption{Two-particle separation distribution widths for the
realistic rescattering model coupled with three initial state model
cases (see text for definitions of these). 
\label{fig:rescrad}}
\end{figure}

\begin{figure}[tb]
\begin{center}
\includegraphics[angle=0,width=0.48 \textwidth]{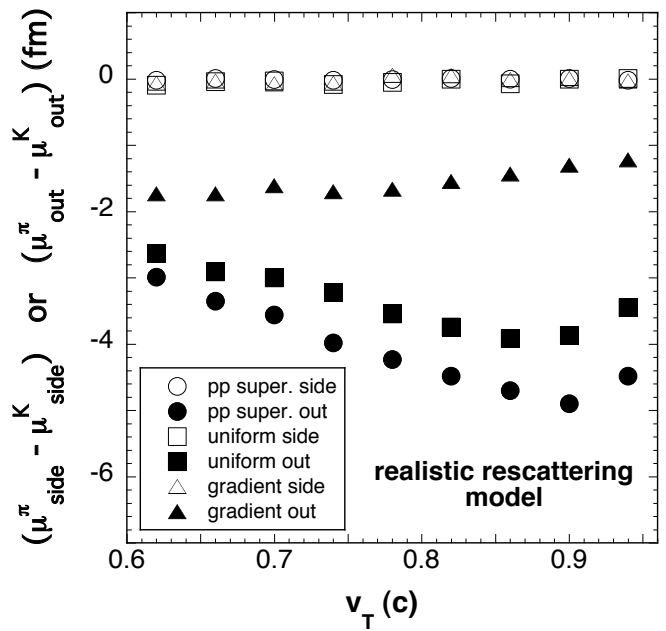}
\end{center}
\vspace{-6.5mm} 
\caption{Mean emission asymmetry between pions and kaons for the
realistic rescattering model coupled with three initial state
model cases (see text for definitions of these).
\label{fig:rescshift}}
\end{figure}

For the results using this initial state model presented in this work,  $\tau$ is set to 0.1 fm/c as in 
Ref. \cite{Humanic:2008nt}. These results are
shown in Figs.~\ref{fig:rescrad} and ~\ref{fig:rescshift}.

In Fig.~\ref{fig:rescrad} we show the two-particle separation
distribution widths for pions and kaons. 
These were extracted by fitting a Gaussian near the peak
of the two-particle separation distributions and extracting the
width, done in the spirit of Ref.~\cite{Hardtke:1999vf} which shows that
the HBT radius parameters are most closely related to the curvature 
of the two-particle space-time relative 
position distribution at the origin.
Thus, these should be comparable
to the ``HBT radii'' obtained from the simple model calculations
above. The conclusions drawn from the simplistic model are valid for
this more realistic simulation. We begin by focusing on the
``gradient'' and ``uniform scenarios, and comparing them to the simple
model. The $m_{T}$ dependence of radii is still steeper for the
gradient case, which means that for the particular initial conditions
the realistic cross-section produce results similar to intermediate
values of the parameter $d$ from the simple model (between $2$ and
$5$~fm). However when one also takes into account the more realistic 
``pp superposition'' model initial size of the
system, the system starts to resemble the simple one with maximum $d$. 

In Fig.~\ref{fig:rescshift} the emission asymmetry between pions and
kaons is shown for all simulation scenarios. 
A method similar to that used to extract the widths in Fig.~\ref{fig:rescrad}
is used to extract the emission asymmetry by fitting a Gaussian near the
peak of the two-particle separation distribution and extracting the peak
position~\cite{Hardtke:1999vf}.
Again the conclusions
from the simplistic model hold, but more can be learned from the realistic
simulation. The asymmetry in the side direction always vanishes, as
required. For the simple initial conditions with realistic
cross-sections we see a departure from the behavior seen in the simple
model. The ``gradient'' case shows a small asymmetry, comparable to
the one observed for the simple model with $d=2$~fm. In the simple model
changing to the ``uniform'' initial conditions did not change the
asymmetry, but with the realistic cross-sections it does, it increases
them by a factor of up to 2. It is also seen that the ``uniform'' case more
closely agrees with the more realistic ``pp superposition'' initial state.

A possible explanation for the difference in emission asymmetry 
between the ``uniform'' and ``gradient'' cases is as follows. 
In the ``uniform'' case, where $T=300$~MeV everywhere, kaons at the
center will have a small velocity, and they will not escape fast. They
will stay with the system longer than the lighter and thus faster
pions, needing  more time to rescatter and build up their transverse
velocity to be in the same $v_T$ bin with pions. As a result their
last interaction point will be more shifted to the outside of  
the source - hence the larger emission asymmetry.
In the ``gradient'' case where $T=500$~MeV at the center where most
of the particles initially reside, both pions and kaons will initially have
higher energy than for the ``uniform'' case. From the rescattering calculation
we find that the $\pi-\pi$ scattering rate to produce $\rho$ resonances increases
over the ``uniform'' case by about 10\%, which moves the average last interaction
point for pions closer to that of the kaons (whose emission points are found
to be less affected for this case then those for pions), reducing the emission 
asymmetry compared
with the ``uniform'' case. 

We also emphasize
that for the more realistic ``pp superposition'' model case the emission 
asymmetry is the largest,
again confirming the any realistic calculation strongly favors the
development of collectivity.

\section{Conclusions}

We used a simple rescattering model to study the behavior of
observables, which were proposed as signatures of collective behavior
of matter. They are of special interest, since collectivity is thought
to arise via many particle interactions, which should also result in
isotropization, and eventually in thermalization of
matter. Alternative scenarios for the development of some collectivity
signatures - the ``temperature gradients'' - were proposed that did
not require thermalization. Since our model was very simple we were
not forced to assume thermalization, and therefore we were able to
simulate both cases: ``collectivity'' and ``gradient'', and show how
they are reflected in all the relevant observables.

We found, in agreement with previous works, that if one only
studies single-particle HBT radii and one does not take into account
the particle mass, one can produce similar $m_{T}$ dependence of radii
with ``gradient'' and ``collective'' scenarios. However as soon as one
considers particle's mass, clear differences arise. Only in the
collective scenario produces a ``universal'' $m_{T}$ scaling for pions
and kaons, as observed in the RHIC data. In addition the pion-kaon
emission asymmetry was shown to be a very clean probe of
collectivity. No interactions (no collective velocity) produces no
asymmetry, while many interactions produces a very significant
one. Moreover it was shown the the initial conditions, whether they
showed gradients in initial temperatures or uniform distributions,
produce similar results, as soon as significant amount of
rescatterings were introduced into the system.

Repeating calculations with the realistic rescattering model
confirmed all of the conclusions. In addition it showed a preference
for an initial thermal model with a uniform temperature distribution
over the system and not for an initial temperature gradient.

Therefore we have identified femtoscopic observables: i.e. the $m_{T}$
dependence of radii for different mass particles and emission
asymmetries between non-identical particles, that are able to cleanly
and unambiguously differentiate between a collective and non-collective
system. We also emphasize that the development of collectivity is
intimately related with themalization, therefore the observation of
one implies at least some degree of the other. All available
experimental data seems to be consistent with the collective
hypothesis, while being in direct contradiction with the lack of
collectivity. We conclude that it is possible, via the femtoscopic
observables to confirm the collective and thermal nature of the system
produced in the heavy ion collisions.

\bibliography{rescatter_ak5}

\begin{thebibliography}{27}
\expandafter\ifx\csname natexlab\endcsname\relax\def\natexlab#1{#1}\fi
\expandafter\ifx\csname bibnamefont\endcsname\relax
  \def\bibnamefont#1{#1}\fi
\expandafter\ifx\csname bibfnamefont\endcsname\relax
  \def\bibfnamefont#1{#1}\fi
\expandafter\ifx\csname citenamefont\endcsname\relax
  \def\citenamefont#1{#1}\fi
\expandafter\ifx\csname url\endcsname\relax
  \def\url#1{\texttt{#1}}\fi
\expandafter\ifx\csname urlprefix\endcsname\relax\def\urlprefix{URL }\fi
\providecommand{\bibinfo}[2]{#2}
\providecommand{\eprint}[2][]{\url{#2}}

\bibitem[{\citenamefont{Abelev et~al.}(2009)}]{Abelev:2008ez}
\bibinfo{author}{\bibfnamefont{B.~I.} \bibnamefont{Abelev}}
  \bibnamefont{et~al.} (\bibinfo{collaboration}{STAR}), \bibinfo{journal}{Phys.
  Rev.} \textbf{\bibinfo{volume}{C79}}, \bibinfo{pages}{034909}
  (\bibinfo{year}{2009}), \eprint{0808.2041}.

\bibitem[{\citenamefont{Voloshin et~al.}(2008)\citenamefont{Voloshin,
  Poskanzer, and Snellings}}]{Voloshin:2008dg}
\bibinfo{author}{\bibfnamefont{S.~A.} \bibnamefont{Voloshin}},
  \bibinfo{author}{\bibfnamefont{A.~M.} \bibnamefont{Poskanzer}},
  \bibnamefont{and} \bibinfo{author}{\bibfnamefont{R.}~\bibnamefont{Snellings}}
  (\bibinfo{year}{2008}), \eprint{0809.2949}.

\bibitem[{\citenamefont{Adams et~al.}(2005)}]{Adams:2004yc}
\bibinfo{author}{\bibfnamefont{J.}~\bibnamefont{Adams}} \bibnamefont{et~al.}
  (\bibinfo{collaboration}{STAR}), \bibinfo{journal}{Phys. Rev.}
  \textbf{\bibinfo{volume}{C71}}, \bibinfo{pages}{044906}
  (\bibinfo{year}{2005}), \eprint{nucl-ex/0411036}.

\bibitem[{\citenamefont{Adams et~al.}(2004)}]{Adams:2003ra}
\bibinfo{author}{\bibfnamefont{J.}~\bibnamefont{Adams}} \bibnamefont{et~al.}
  (\bibinfo{collaboration}{STAR}), \bibinfo{journal}{Phys. Rev. Lett.}
  \textbf{\bibinfo{volume}{93}}, \bibinfo{pages}{012301}
  (\bibinfo{year}{2004}), \eprint{nucl-ex/0312009}.

\bibitem[{\citenamefont{Adams et~al.}(2003)}]{Adams:2003qa}
\bibinfo{author}{\bibfnamefont{J.}~\bibnamefont{Adams}} \bibnamefont{et~al.}
  (\bibinfo{collaboration}{STAR}), \bibinfo{journal}{Phys. Rev. Lett.}
  \textbf{\bibinfo{volume}{91}}, \bibinfo{pages}{262302}
  (\bibinfo{year}{2003}), \eprint{nucl-ex/0307025}.

\bibitem[{\citenamefont{Hirano and Tsuda}(2002)}]{Hirano:2002ds}
\bibinfo{author}{\bibfnamefont{T.}~\bibnamefont{Hirano}} \bibnamefont{and}
  \bibinfo{author}{\bibfnamefont{K.}~\bibnamefont{Tsuda}},
  \bibinfo{journal}{Phys. Rev.} \textbf{\bibinfo{volume}{C66}},
  \bibinfo{pages}{054905} (\bibinfo{year}{2002}), \eprint{nucl-th/0205043}.

\bibitem[{\citenamefont{Zschiesche et~al.}(2002)\citenamefont{Zschiesche,
  Schramm, Stoecker, and Greiner}}]{Zschiesche:2001dx}
\bibinfo{author}{\bibfnamefont{D.}~\bibnamefont{Zschiesche}},
  \bibinfo{author}{\bibfnamefont{S.}~\bibnamefont{Schramm}},
  \bibinfo{author}{\bibfnamefont{H.}~\bibnamefont{Stoecker}}, \bibnamefont{and}
  \bibinfo{author}{\bibfnamefont{W.}~\bibnamefont{Greiner}},
  \bibinfo{journal}{Phys. Rev.} \textbf{\bibinfo{volume}{C65}},
  \bibinfo{pages}{064902} (\bibinfo{year}{2002}), \eprint{nucl-th/0107037}.

\bibitem[{\citenamefont{Heinz and Kolb}(2002)}]{Heinz:2002un}
\bibinfo{author}{\bibfnamefont{U.~W.} \bibnamefont{Heinz}} \bibnamefont{and}
  \bibinfo{author}{\bibfnamefont{P.~F.} \bibnamefont{Kolb}}
  (\bibinfo{year}{2002}), \eprint{hep-ph/0204061}.

\bibitem[{\citenamefont{Broniowski et~al.}(2008)\citenamefont{Broniowski,
  Chojnacki, Florkowski, and Kisiel}}]{Broniowski:2008vp}
\bibinfo{author}{\bibfnamefont{W.}~\bibnamefont{Broniowski}},
  \bibinfo{author}{\bibfnamefont{M.}~\bibnamefont{Chojnacki}},
  \bibinfo{author}{\bibfnamefont{W.}~\bibnamefont{Florkowski}},
  \bibnamefont{and} \bibinfo{author}{\bibfnamefont{A.}~\bibnamefont{Kisiel}},
  \bibinfo{journal}{Phys. Rev. Lett.} \textbf{\bibinfo{volume}{101}},
  \bibinfo{pages}{022301} (\bibinfo{year}{2008}), \eprint{0801.4361}.

\bibitem[{\citenamefont{Pratt}(2009)}]{Pratt:2008qv}
\bibinfo{author}{\bibfnamefont{S.}~\bibnamefont{Pratt}},
  \bibinfo{journal}{Phys. Rev. Lett.} \textbf{\bibinfo{volume}{102}},
  \bibinfo{pages}{232301} (\bibinfo{year}{2009}), \eprint{0811.3363}.

\bibitem[{\citenamefont{Gombeaud et~al.}(2009)\citenamefont{Gombeaud, Lappi,
  and Ollitrault}}]{Gombeaud:2009fk}
\bibinfo{author}{\bibfnamefont{C.}~\bibnamefont{Gombeaud}},
  \bibinfo{author}{\bibfnamefont{T.}~\bibnamefont{Lappi}}, \bibnamefont{and}
  \bibinfo{author}{\bibfnamefont{J.-Y.} \bibnamefont{Ollitrault}},
  \bibinfo{journal}{Phys. Rev.} \textbf{\bibinfo{volume}{C79}},
  \bibinfo{pages}{054914} (\bibinfo{year}{2009}), \eprint{0901.4908}.

\bibitem[{\citenamefont{Zhang et~al.}(1998)\citenamefont{Zhang, Gyulassy, and
  Pang}}]{Zhang:1998tj}
\bibinfo{author}{\bibfnamefont{B.}~\bibnamefont{Zhang}},
  \bibinfo{author}{\bibfnamefont{M.}~\bibnamefont{Gyulassy}}, \bibnamefont{and}
  \bibinfo{author}{\bibfnamefont{Y.}~\bibnamefont{Pang}},
  \bibinfo{journal}{Phys. Rev.} \textbf{\bibinfo{volume}{C58}},
  \bibinfo{pages}{1175} (\bibinfo{year}{1998}), \eprint{nucl-th/9801037}.

\bibitem[{\citenamefont{Cheng et~al.}(2002)}]{Cheng:2001dz}
\bibinfo{author}{\bibfnamefont{S.}~\bibnamefont{Cheng}} \bibnamefont{et~al.},
  \bibinfo{journal}{Phys. Rev.} \textbf{\bibinfo{volume}{C65}},
  \bibinfo{pages}{024901} (\bibinfo{year}{2002}), \eprint{nucl-th/0107001}.

\bibitem[{\citenamefont{Molnar and Gyulassy}(2004)}]{Molnar:2002bz}
\bibinfo{author}{\bibfnamefont{D.}~\bibnamefont{Molnar}} \bibnamefont{and}
  \bibinfo{author}{\bibfnamefont{M.}~\bibnamefont{Gyulassy}},
  \bibinfo{journal}{Phys. Rev. Lett.} \textbf{\bibinfo{volume}{92}},
  \bibinfo{pages}{052301} (\bibinfo{year}{2004}), \eprint{nucl-th/0211017}.

\bibitem[{\citenamefont{Retiere and Lisa}(2004)}]{Retiere:2003kf}
\bibinfo{author}{\bibfnamefont{F.}~\bibnamefont{Retiere}} \bibnamefont{and}
  \bibinfo{author}{\bibfnamefont{M.~A.} \bibnamefont{Lisa}},
  \bibinfo{journal}{Phys. Rev.} \textbf{\bibinfo{volume}{C70}},
  \bibinfo{pages}{044907} (\bibinfo{year}{2004}), \eprint{nucl-th/0312024}.

\bibitem[{\citenamefont{Kisiel et~al.}(2006)\citenamefont{Kisiel, Florkowski,
  and Broniowski}}]{Kisiel:2006is}
\bibinfo{author}{\bibfnamefont{A.}~\bibnamefont{Kisiel}},
  \bibinfo{author}{\bibfnamefont{W.}~\bibnamefont{Florkowski}},
  \bibnamefont{and}
  \bibinfo{author}{\bibfnamefont{W.}~\bibnamefont{Broniowski}},
  \bibinfo{journal}{Phys. Rev.} \textbf{\bibinfo{volume}{C73}},
  \bibinfo{pages}{064902} (\bibinfo{year}{2006}), \eprint{nucl-th/0602039}.

\bibitem[{\citenamefont{Chojnacki and Florkowski}(2006)}]{Chojnacki:2006tv}
\bibinfo{author}{\bibfnamefont{M.}~\bibnamefont{Chojnacki}} \bibnamefont{and}
  \bibinfo{author}{\bibfnamefont{W.}~\bibnamefont{Florkowski}},
  \bibinfo{journal}{Phys. Rev.} \textbf{\bibinfo{volume}{C74}},
  \bibinfo{pages}{034905} (\bibinfo{year}{2006}), \eprint{nucl-th/0603065}.

\bibitem[{\citenamefont{Frodermann et~al.}(2006)\citenamefont{Frodermann,
  Heinz, and Lisa}}]{Frodermann:2006sp}
\bibinfo{author}{\bibfnamefont{E.}~\bibnamefont{Frodermann}},
  \bibinfo{author}{\bibfnamefont{U.}~\bibnamefont{Heinz}}, \bibnamefont{and}
  \bibinfo{author}{\bibfnamefont{M.~A.} \bibnamefont{Lisa}},
  \bibinfo{journal}{Phys. Rev.} \textbf{\bibinfo{volume}{C73}},
  \bibinfo{pages}{044908} (\bibinfo{year}{2006}), \eprint{nucl-th/0602023}.

\bibitem[{\citenamefont{Lednicky et~al.}(1996)\citenamefont{Lednicky,
  Lyuboshits, Erazmus, and Nouais}}]{Lednicky:1995vk}
\bibinfo{author}{\bibfnamefont{R.}~\bibnamefont{Lednicky}},
  \bibinfo{author}{\bibfnamefont{V.~L.} \bibnamefont{Lyuboshits}},
  \bibinfo{author}{\bibfnamefont{B.}~\bibnamefont{Erazmus}}, \bibnamefont{and}
  \bibinfo{author}{\bibfnamefont{D.}~\bibnamefont{Nouais}},
  \bibinfo{journal}{Phys. Lett.} \textbf{\bibinfo{volume}{B373}},
  \bibinfo{pages}{30} (\bibinfo{year}{1996}).

\bibitem[{\citenamefont{Lednicky et~al.}(2003)\citenamefont{Lednicky, Panitkin,
  and Xu}}]{Lednicky:2003fe}
\bibinfo{author}{\bibfnamefont{R.}~\bibnamefont{Lednicky}},
  \bibinfo{author}{\bibfnamefont{S.}~\bibnamefont{Panitkin}}, \bibnamefont{and}
  \bibinfo{author}{\bibfnamefont{N.}~\bibnamefont{Xu}} (\bibinfo{year}{2003}),
  \eprint{nucl-th/0304062}.

\bibitem[{\citenamefont{Kisiel}(2009)}]{Kisiel:2009zz}
\bibinfo{author}{\bibfnamefont{A.}~\bibnamefont{Kisiel}},
  \bibinfo{journal}{Acta Phys. Polon.} \textbf{\bibinfo{volume}{B40}},
  \bibinfo{pages}{1155} (\bibinfo{year}{2009}).

\bibitem[{\citenamefont{Humanic}(2006)}]{Humanic:2005ye}
\bibinfo{author}{\bibfnamefont{T.~J.} \bibnamefont{Humanic}},
  \bibinfo{journal}{Int. J. Mod. Phys.} \textbf{\bibinfo{volume}{E15}},
  \bibinfo{pages}{197} (\bibinfo{year}{2006}), \eprint{nucl-th/0510049}.

\bibitem[{\citenamefont{Prakash et~al.}(1993)\citenamefont{Prakash, Prakash,
  Venugopalan, and Welke}}]{Prakash:1993bt}
\bibinfo{author}{\bibfnamefont{M.}~\bibnamefont{Prakash}},
  \bibinfo{author}{\bibfnamefont{M.}~\bibnamefont{Prakash}},
  \bibinfo{author}{\bibfnamefont{R.}~\bibnamefont{Venugopalan}},
  \bibnamefont{and} \bibinfo{author}{\bibfnamefont{G.}~\bibnamefont{Welke}},
  \bibinfo{journal}{Phys. Rept.} \textbf{\bibinfo{volume}{227}},
  \bibinfo{pages}{321} (\bibinfo{year}{1993}).

\bibitem[{\citenamefont{Yao et~al.}(2006)}]{Yao:2006px}
\bibinfo{author}{\bibfnamefont{W.~M.} \bibnamefont{Yao}} \bibnamefont{et~al.}
  (\bibinfo{collaboration}{Particle Data Group}), \bibinfo{journal}{J. Phys.}
  \textbf{\bibinfo{volume}{G33}}, \bibinfo{pages}{1} (\bibinfo{year}{2006}).

\bibitem[{\citenamefont{Humanic}(2009)}]{Humanic:2008nt}
\bibinfo{author}{\bibfnamefont{T.~J.} \bibnamefont{Humanic}},
  \bibinfo{journal}{Phys. Rev.} \textbf{\bibinfo{volume}{C79}},
  \bibinfo{pages}{044902} (\bibinfo{year}{2009}), \eprint{0810.0621}.

\bibitem[{\citenamefont{Sjostrand et~al.}(2006)\citenamefont{Sjostrand, Mrenna,
  and Skands}}]{Sjostrand:2006za}
\bibinfo{author}{\bibfnamefont{T.}~\bibnamefont{Sjostrand}},
  \bibinfo{author}{\bibfnamefont{S.}~\bibnamefont{Mrenna}}, \bibnamefont{and}
  \bibinfo{author}{\bibfnamefont{P.}~\bibnamefont{Skands}},
  \bibinfo{journal}{JHEP} \textbf{\bibinfo{volume}{05}}, \bibinfo{pages}{026}
  (\bibinfo{year}{2006}), \eprint{hep-ph/0603175}.

\bibitem[{\citenamefont{Hardtke and Voloshin}(2000)}]{Hardtke:1999vf}
\bibinfo{author}{\bibfnamefont{D.}~\bibnamefont{Hardtke}} \bibnamefont{and}
  \bibinfo{author}{\bibfnamefont{S.~A.} \bibnamefont{Voloshin}},
  \bibinfo{journal}{Phys. Rev.} \textbf{\bibinfo{volume}{C61}},
  \bibinfo{pages}{024905} (\bibinfo{year}{2000}), \eprint{nucl-th/9906033}.

\end{thebibliography}

\end{document}